\documentclass[aps,prd,twocolumn,floats,floatfix,letterpaper,nofootinbib,preprintnumbers,superscriptaddress]{revtex4-1}

\usepackage{amssymb,amsmath,latexsym,mathrsfs}
\usepackage{graphicx}
\usepackage{epsfig}
\usepackage{multirow}
\usepackage{array}
\usepackage{xspace}
\usepackage{color,xcolor}
\usepackage{physics}
\usepackage{booktabs}
\usepackage[normalem]{ulem}

\def\linkcolor{cyan!70!black}
\usepackage[
colorlinks=true
,urlcolor=\linkcolor
,anchorcolor=\linkcolor
,citecolor=\linkcolor
,filecolor=\linkcolor
,linkcolor=\linkcolor
,menucolor=\linkcolor
,linktocpage=true
,pdfproducer=medialab
,pdfa=true
]{hyperref}

\usepackage{epsfig,psfrag,rotating,soul}
\usepackage{rotfloat}
\usepackage{slashed}

\makeatletter
\newcommand{\mathleft}{\@fleqntrue\@mathmargin0pt}
\newcommand{\mathcenter}{\@fleqnfalse}
\makeatother

\allowdisplaybreaks

\begin{document}


\title{\vspace*{.5cm}
Indirect upper limits on $\boldsymbol{\ell_i\to\ell_j\gamma\gamma}$ from $\boldsymbol{\ell_i\to\ell_j\gamma}$}

\author{Fabiola Fortuna}
\affiliation{Departamento de F\'isica, Centro de Investigaci\'on y de Estudios Avanzados del Instituto Polit\'ecnico Nacional, Apdo. Postal 14-740, 07000 Ciudad de M\'exico, M\'exico.}

\author{Alejandro Ibarra}
\affiliation{Physik-Department, Technische Universit\"at M\"unchen, James-Franck-Stra\ss e, 85748 Garching, Germany}

\author{Xabier Marcano}
\affiliation{Departamento de F\'{\i}sica Te\'orica and Instituto de F\'{\i}sica Te\'orica UAM/CSIC,\\
Universidad Aut\'onoma de Madrid, Cantoblanco, 28049 Madrid, Spain}
\author{Marcela Mar\'in}
\affiliation{Departamento de F\'isica, Centro de Investigaci\'on y de Estudios Avanzados del Instituto Polit\'ecnico Nacional, Apdo. Postal 14-740, 07000 Ciudad de M\'exico, M\'exico.}
\author{Pablo Roig}
\affiliation{Departamento de F\'isica, Centro de Investigaci\'on y de Estudios Avanzados del Instituto Polit\'ecnico Nacional, Apdo. Postal 14-740, 07000 Ciudad de M\'exico, M\'exico.}

\begin{abstract}

We perform an effective field theory analysis to correlate the charged lepton flavor violating processes $\ell_i\to\ell_j\gamma\gamma$ and $\ell_i\to\ell_j\gamma$. Using the current upper bounds on the rate for  $\ell_i\to\ell_j\gamma$, we derive model-independent upper limits on the rates for $\ell_i\to\ell_j\gamma\gamma$. Our indirect limits are about three orders of magnitude stronger than the direct bounds from current searches for $\mu\to e\gamma\gamma$, and four orders of magnitude better than current bounds for $\tau\to\ell\gamma\gamma$. We also stress the relevance of Belle II or a Super Tau Charm Facility to discover the rare decay   $\tau\to\ell\gamma\gamma$.
\end{abstract}

\preprint{TUM-HEP 1422/22}
\preprint{IFT-UAM/CSIC-22-124}

\maketitle


\section{Introduction}\label{sec:Intro}

\begin{figure}[t!]
\begin{center}
\includegraphics[width=.95\columnwidth]{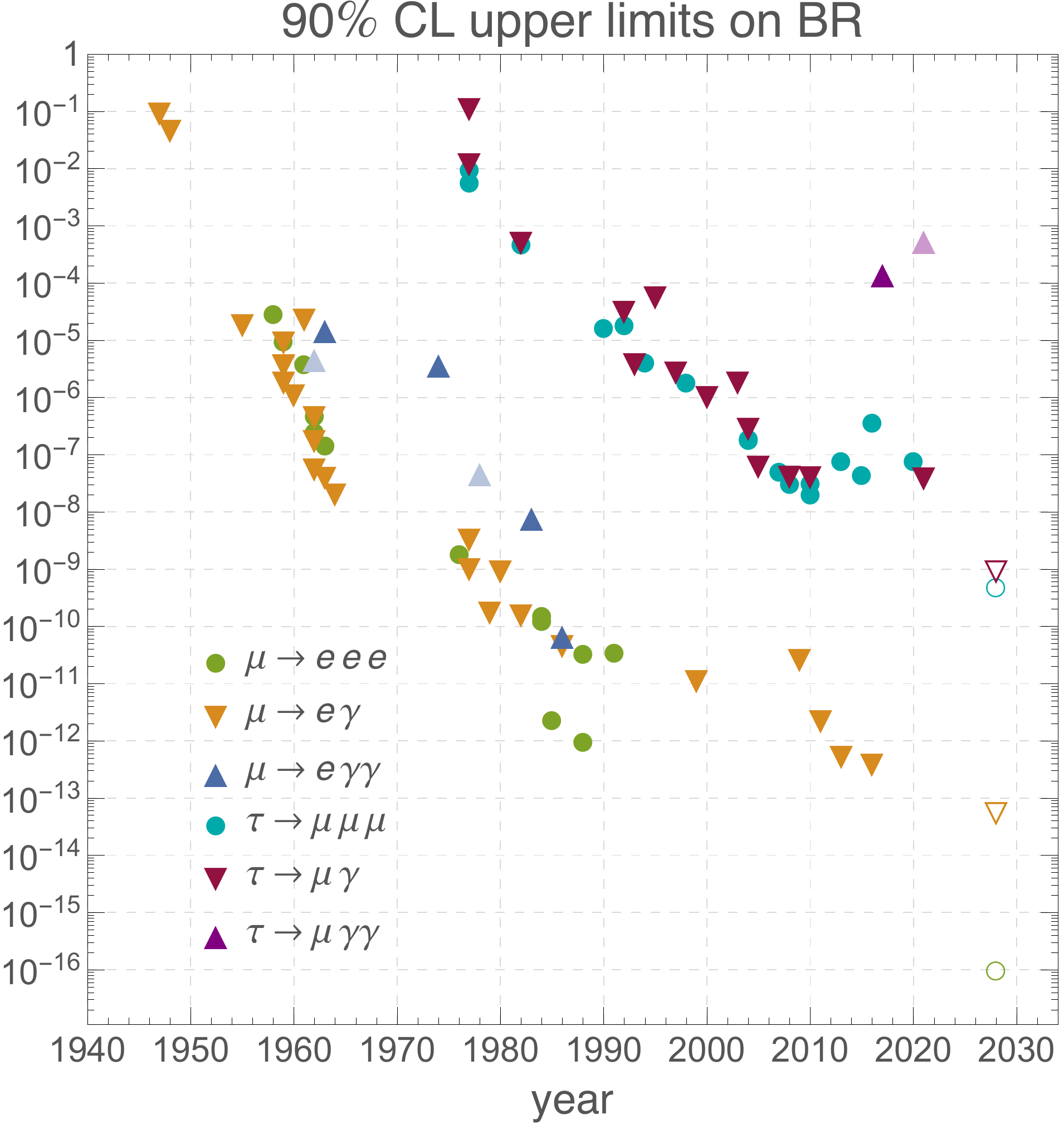}
\caption{Historical evolution for the 90\%CL upper limits on several cLFV leptonic decays.
Solid markers correspond to published direct experimental searches, while empty ones indicate future expected sensitivities at MEG II~\cite{MEGII:2018kmf}, Mu3e~\cite{Blondel:2013ia} and Belle II~\cite{Belle-II:2018jsg} (arbitrary year).
Lighter markers for $\ell_i\to\ell_j\gamma\gamma$ were obtained by recasting the available searches for $\ell_i\to\ell_j\gamma$~\cite{Bowman:1978kz,Bryman:2021ilc} or $\mu e \to \gamma\gamma$~\cite{PhysRev.126.375}. 
The evolution for $\tau\to e$ sector is similar to the $\tau\to \mu$ one, with the exception that no direct search for $\tau\to e\gamma \gamma$ exists.
}\label{LFVhistoryplot}
\end{center}
\end{figure}

The experimental observation of charged lepton flavor violation (cLFV) would undoubtedly imply the existence of new physics beyond neutrino oscillations~\cite{Calibbi:2017uvl}. 
This has motivated a strong experimental program over the last 75 years searching for different cLFV processes, see Fig.~\ref{LFVhistoryplot}, each observable providing complementary information about possible beyond the Standard Model (BSM) scenarios. 

Here we consider the cLFV decays of leptons to two photons, $\ell_i\to\ell_j\gamma\gamma$~\cite{PhysRev.126.375,Bowman:1978kz, Davidson:2020ord}, which have been explored in less detail than other cLFV processes such as the single photon process,  $\ell_i\to\ell_j\gamma$, specially for the case of $\tau\to\ell\gamma\gamma$ \cite{Gemintern:2003gd,Cordero-Cid:2005vca, Aranda:2008si, Aranda:2009kz,Bryman:2021ilc}.

Experimentally, $\mu\to e\gamma\gamma$ was searched for by several experiments aiming also for $\mu\to e\gamma$. 
The latest of these experiments was the Crystal Box detector, whose result still provides the strongest bound for $\mu\to e\gamma\gamma$~\cite{Grosnick:1986pr}.
This limit is however two orders of magnitude weaker than present $\mu\to e\gamma$ bounds, see Table~\ref{tab:BRsExp}, since the MEG experiment was optimized for back-to-back topologies and no new dedicated experiment for $\mu\to e\gamma\gamma$ has been carried out since Crystal Box. 
On the other hand, $\tau\to\ell\gamma\gamma$ has rarely been searched for. 
To the best of our knowledge, the only existing direct experimental search was performed by ATLAS, setting an upper limit of BR$(\tau\to\mu\gamma\gamma)<1.5\times10^{-4}$ after the LHC run-I~\cite{Angelozzi:2017oeg}.  
No direct experimental search exists for $\tau\to e\gamma\gamma$. 

An alternative for exploring the $\ell_i\to\ell_j\gamma\gamma$ channels is to recast the searches for $\ell_i\to\ell_j\gamma$, as some of the events of the former would fall into the signal region defined for the latter~\cite{Bowman:1978kz}.
This idea has been recently applied to recast the BABAR search for $\tau\to\ell\gamma$~\cite{Aubert:2009ag}, finding that at 90\%CL BR$(\tau\to\mu\gamma\gamma)<5.8\times10^{-4}$ and BR$(\tau\to e\gamma\gamma)<2.5\times10^{-4}$~\cite{Bryman:2021ilc}.
These limits are however several orders of magnitude weaker than the associated ones on $\tau\to\ell\gamma$ due to the low acceptance of these searches for $\tau\to\ell\gamma\gamma$ events. 

In this letter, we consider the theoretical correlation between the $\ell_i\to\ell_j\gamma\gamma$ and the $\ell_i\to\ell_j\gamma$ decays. Clearly, any scenario generating $\ell_i\to\ell_j\gamma$ would automatically generate a (model-independent) contribution to $\ell_i\to\ell_j\gamma\gamma$, from the  radiation of an additional photon in the final state.  Further, any scenario generating $\ell_i\to\ell_j\gamma\gamma$ will generate a (model-dependent) contribution to $\ell_i\to\ell_j\gamma$ at the quantum level. Barring cancellations, the quantum-induced contribution should not exceed the experimental upper limits on  $\ell_i\to\ell_j\gamma$, which in turn allows to set indirect limits on the rates of $\ell_i\to\ell_j\gamma\gamma$. 

In this work, we will pursue an effective field theory (EFT) approach to study these correlations, in order to ensure the generality of our conclusions. Notably, our indirect limits on the process $\ell_i\to\ell_j\gamma\gamma$ will turn out to be more stringent than the current direct bounds. Furthermore, these limits do not preclude the possibility of observing the rare decays $\tau\to\ell\gamma\gamma$ at Belle II or at a Super Tau Charm Factory, which then represents a competitive probe of cLFV along with the more studied channels $\tau\to\ell\gamma$ or $\tau\to \ell_j \ell_k \bar \ell_k $.

\begin{table}[t!]
\begin{center}
\renewcommand{\arraystretch}{1.3}
\setlength{\tabcolsep}{8pt}
\begin{tabular}{lll}
\hline
\hline
Decay Mode & \multicolumn{2}{l}{Current upper limit on BR (90\%CL)}\\
\hline
$\mu\to e\gamma$ &  $4.2\times10^{-13}$ & MEG (2016)~\cite{TheMEG:2016wtm} \\
$\mu\to e\gamma\gamma$ &  $7.2\times10^{-11}$ & Crystal Box (1986)~\cite{Grosnick:1986pr} \\
$\tau\to e\gamma$ &  $3.3\times10^{-8}$ &BaBar (2010)~\cite{Aubert:2009ag} \\
$\tau\to \mu\gamma$ &  $4. 2\times10^{-8}$ &Belle (2021)~\cite{Belle:2021ysv} \\
$\tau\to \mu\gamma\gamma$ &  $1.5\times10^{-4}$ & ATLAS (2017)~\cite{Angelozzi:2017oeg} \\
\hline
\hline
\end{tabular}
 \caption{Experimental upper bounds on the rates of the $\ell_i\to\ell_j\gamma(\gamma)$ decays.} \label{tab:BRsExp}  

\end{center}
\end{table}

\section{Total rate for $\boldsymbol{\ell_i\to\ell_j\gamma\gamma}$ in the EFT approach}\label{sec:EFT2photon}

The effective interaction Lagrangian between two charged leptons of different flavor and one photon has dimension 5 and reads:
\begin{equation}\label{eq:Leffdim5}
\mathcal{L}_\text{dim-5} = 
D_{R}^{ij}\, \bar{\ell}_{L_i} \sigma_{\mu\nu} \ell_{R_j} F^{\mu\nu} 
+ D_{L}^{ij}\, \bar{\ell}_{R_i} \sigma_{\mu\nu} \ell_{L_j} F^{\mu\nu}
+h.c.
\end{equation}
where the subscripts $L(R)$ indicate the chirality of the lepton  and $i,j$ are generation indices. 

On the other hand, the lowest dimensional effective interaction between two charged leptons of different flavor and two photons has mass dimension 7 and is given by~\cite{Bowman:1978kz}
\mathleft
\begin{align} \label{eq:Leff}
  \mathcal{L}_\text{dim-7}=&\left(G_{SR}^{\, ij}\bar{\ell}_{L_i}\ell_{R_j}+G_{SL}^{\, ij}\bar{\ell}_{R_i}\ell_{L_j}\right)F_{\mu\nu} F^{\mu\nu}\nonumber\\
 +&\left(\tilde{G}_{SR}^{\,ij}\bar{\ell}_{L_i}\ell_{R_j}+\tilde{G}_{SL}^{\,ij}\bar{\ell}_{R_i}\ell_{L_j}\right)\tilde{F}_{\mu\nu}F^{\mu\nu} +
 h. c.
\end{align}
\mathcenter
where  $\tilde{F}_{\mu\nu}=\frac{1}{2}\epsilon_{\mu\nu\sigma\lambda}F^{\sigma\lambda}$ is the dual tensor.  
In Ref.~\cite{Bowman:1978kz}, the Lagrangian also contains the dimension-8 operators $\bar{\ell}_{L_i}\gamma^\sigma\ell_{L_j}F^{\mu\nu}\partial_\nu F_{\mu\sigma}$ and $\bar{\ell}_{L_i}\gamma^\sigma\ell_{L_j}F^{\mu\nu}\partial_\nu \tilde{F}_{\mu\sigma}$, as well as the analogous operators for the right-handed fermions. The effect of these operators is not only suppressed by higher powers of the cut-off scale of the EFT, but also by the mass of the decaying lepton, due to the helicity flip, therefore we will neglect them henceforth.

The effective Lagrangians in Eqs.~\eqref{eq:Leffdim5} and \eqref{eq:Leff} generate at tree level the decay $\ell_i\to\ell_j\gamma\gamma$, through the diagrams shown in Fig.~\ref{fig:diagrams}. The expression for the differential decay rate is  complicated and is given in  Appendix~\ref{sec:Appendix}. For the specific limits where the rate is dominated by the dimension-5 operators, the total decay rate is given by:
\begin{equation}\label{eq:rate_2gamma_d5}
\Gamma(\ell_i\to\ell_j\gamma\gamma) =  \frac{\alpha\, m_i^3 }{48 \pi ^2} \Big(|D^{ij}_R|^2 + |D^{ij}_L|^2\Big)\,
\lambda\left(\tfrac{E_\gamma^{\rm cut}}{m_i}\right) \,,
\end{equation}
with $E_\gamma^{\rm cut}$ an energy cut-off introduced to regularize the infrared and collinear divergences in the rate (see App.~\ref{sec:Appendix}), and 
\begin{align}
\lambda(x)&\simeq  6 + 2\pi^2+6\log^22 
+21\log(2x)+6\log(x)\,\log(4x)\nonumber\\
& +18 x \big(2\log(2x)+1\big)
+ 6x^2\big(8\log(2x)-29\big) \nonumber\\
&+\mathcal O(x^3)\,.
\end{align}
On the other hand, when it is dominated by the dimension-7 operators, we obtain
\begin{equation} \label{eq:rate_2gamma}
\Gamma(\ell_i\to\ell_j\gamma\gamma) =  \frac{|G_{ij}|^2 }{3840 \pi ^3}\, m_i^7 \,,
\end{equation}
where we have neglected the mass of the final lepton and  $|G_{ij}|^2 = |G_{SL}^{\,ij}|^2 + |G_{SR}^{\,ij}|^2 + |\tilde G_{SL}^{\,ij}|^2 + |\tilde G_{SR}^{\,ij}|^2$.

\begin{figure*}[t!]
 \centering
\includegraphics[width=.9\textwidth]{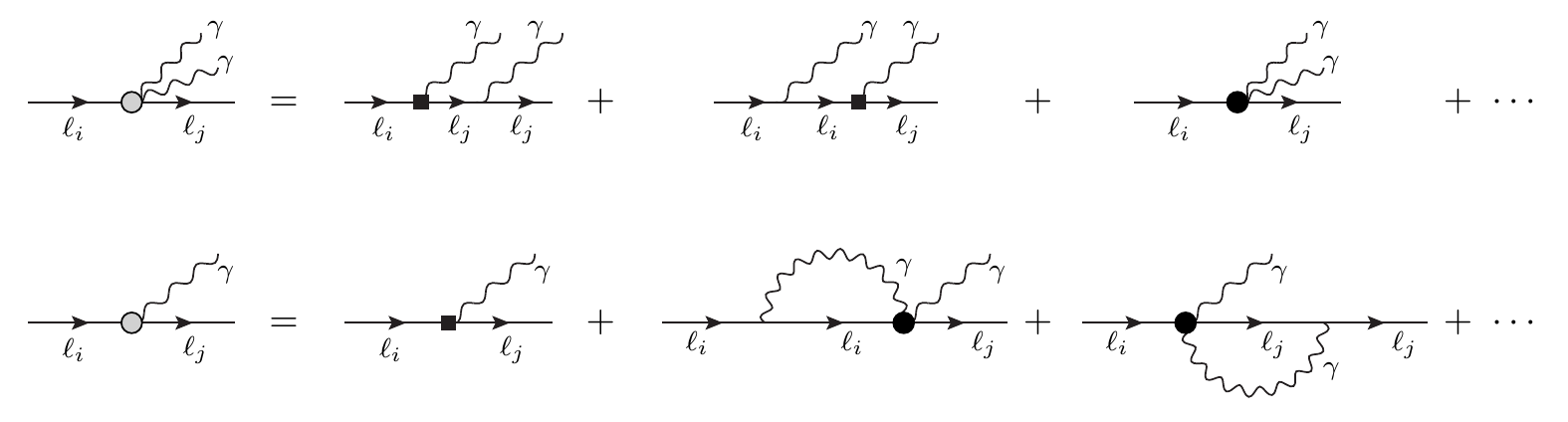} 
\caption{Feynman diagrams contributing to $\ell_i \to \ell_j \gamma\gamma$ (top panel) and $\ell_i \to \ell_j \gamma$ (bottom panel) in the effective field theory described by the Lagrangians in Eq.~\eqref{eq:Leffdim5} (black squares) and Eq.~\eqref{eq:Leff} (black circles). The dots represent higher order contributions.}
\label{fig:diagrams}
\end{figure*}

\section{Correlating  $\boldsymbol{\ell_i\to\ell_j\gamma\gamma}$ and $\boldsymbol{\ell_i\to\ell_j\gamma}$}\label{sec:Limits1photon}

Let us first consider scenarios where the dimension-5 operators are not suppressed, so the rate for $\ell_i\rightarrow \ell_j \gamma\gamma$ is approximately given by Eq.~\eqref{eq:rate_2gamma_d5}. Clearly, the dimension-5 operators also induce the decay $\ell_i\rightarrow \ell_j \gamma$ (see lower panel of Fig.~\ref{fig:diagrams}). The rate is given by
\begin{equation}\label{eq:rate_1gamma_d5}
	\Gamma(\ell_i \to  \ell_j \gamma)= \frac{m_i^3}{4 \pi }\left(\big|D_{R}^{ij}\big|^2 + \big|D_{L}^{ij}\big|^2\right)   \,,
\end{equation}
whence one obtains
\begin{equation}  \label{BRrelations_d5}
\Gamma(\ell_i \to \ell_j \gamma\gamma) = \frac{\alpha}{12\pi}\,\lambda\left(\frac{E_\gamma^{\rm cut}}{m_i}\right) \, 
\Gamma(\ell_i \to \ell_j \gamma)
\,.
\end{equation}
Using the upper limits on the rates for  $\ell_i\to\ell_j\gamma$  from Table \ref{tab:BRsExp} and imposing $E_\gamma^{\rm cut}=7\, (50)$~MeV for $\mu\, (\tau)$ decays, one finds the indirect  limits:
\begin{align} 
	&{\rm BR}(\mu\to e\gamma\gamma) \lesssim 2 \times 10^{-16}\,, \nonumber \\ 
	&{\rm BR}(\tau\to e\gamma\gamma) \lesssim 8\times 10^{-11}\,, \nonumber\\ 
	&{\rm BR}(\tau\to \mu\gamma\gamma) \lesssim 1 \times 10^{-10}\,.\label{BRmaxTAUMUbis}
\end{align}

Alternatively,  there could be scenarios where the dimension-5 operators are suppressed, while not the dimension-7 operators (see Section \ref{sec:models}). In this case, the decay $\ell_i\rightarrow \ell_j \gamma$  is induced at the one loop-level (see lower panel of Fig.~\ref{fig:diagrams}).
In this case, and 
keeping only the leading terms, one finds:
\begin{equation} \label{llgamma1loop}
\Gamma(\ell_i \to \ell_j \gamma)\sim \frac{\alpha\, |G_{ij}|^2}{256\, \pi^4}\, m_i^7\, \log^2\left(\frac{\Lambda^2}{m_i^2}\right)\,,
\end{equation}
where $\Lambda$ is the cut-off energy scale of the effective field theory. Using Eq.~\eqref{eq:rate_2gamma} one obtains an approximate correlation between rates
\begin{equation}  \label{BRrelations}
\Gamma(\ell_i \to \ell_j \gamma)\sim \frac{15\alpha}{\pi}\,\log^2\left(\frac{\Lambda^2}{m_i^2}\right)\, \Gamma(\ell_i \to \ell_j \gamma\gamma)
\,,
\end{equation}
from where one can derive  indirect upper limits for  $\ell_i\to\ell_j\gamma\gamma$ from the upper limits on  $\ell_i\to\ell_j\gamma$: 
\mathleft
\begin{align} 
	{\rm BR}(\mu\to e\gamma\gamma) &  \lesssim  6.4 \times 10^{-14}\Big[1+0.15 \log\tfrac{\Lambda}{100\,{\rm GeV}}\Big]^{-2}\hspace*{-.3cm} , \nonumber\\
	{\rm BR}(\tau\to e\gamma\gamma) &\lesssim1.5 \times 10^{-8}\,\, \Big[1+0.25 \log\tfrac{\Lambda}{100\,{\rm GeV}}\Big]^{-2}\hspace*{-.3cm} ,\nonumber\\
	{\rm BR}(\tau\to \mu\gamma\gamma)  & \lesssim 1.9 \times 10^{-8}\,\, \Big[1+0.25 \log\tfrac{\Lambda}{100\,{\rm GeV}}\Big]^{-2}\hspace*{-.38cm} ,
	\label{eq:limits_dim7_dominance}
\end{align}
\mathcenter
which have a mild sensitivity to the cut-off scale. 

Regardless of the underlying physics generating the process $\ell_i\to\ell_j\gamma\gamma$, our indirect limits are significantly more stringent than the current direct limits. Concretely, the limit on $\mu\to e\gamma\gamma$ is about three orders of magnitude stronger than the direct search using the Crystal Box detector and the limit on $\tau\to \mu\gamma\gamma$ is about four orders of magnitude stronger than the direct search performed at ATLAS. Future foreseeable sensitivities of MEG II searching for $\mu\to e\gamma$ and of Belle II for $\tau\to\ell\gamma$ will improve our indirect limits by about one order of magnitude.

Furthermore, the results in Eqs.~\eqref{eq:limits_dim7_dominance} motivate a dedicated experimental search for the  $\tau\to\ell\gamma\gamma$ decays, since this decay might be at the reach of future experiments~\cite{Banerjee:2022xuw}.  
Let us consider the specific case of the Belle~II experiment. Assuming that Belle~II could achieve the same sensitivity for double than for single photon processes, as occurred in the Crystal Box Detector for muon decays~\cite{Bolton:1988af},
Belle~II could probe the $\tau\to\ell\gamma\gamma$ decays with branching ratios as small as $\mathcal O(10^{-9})$~\cite{Banerjee:2022xuw}. If this sensitivity is reached, Belle~II will probe uncharted parameter space of the dimension-7 operators, and possibly find evidence for cLFV by the observation of the decay $\tau\rightarrow \ell\gamma\gamma$.

\section{Models with enhanced $\boldsymbol{\ell_i\to\ell_j\gamma\gamma}$}
\label{sec:models}

There are several scenarios where the dimension-5 operators could be suppressed with respect to the dimension-7 operators, thereby enhancing the rate of $\ell_i\to\ell_j\gamma\gamma$ compared to $\ell_i\to\ell_j\gamma$. For instance, it was argued in~\cite{Bowman:1978kz} that in models where cLFV was mediated by charged heavy leptons~\cite{Wilczek:1977wb}, the GIM suppression~\cite{PhysRevD.2.1285} could be stronger for $\ell_i\to\ell_j\gamma$ than for $\ell_i\to\ell_j\gamma\gamma$.
Also, in Ref.~\cite{Davidson:2020ord} it was argued that in some new physics models the dimension-7 operators could actually arise at $\mathcal O(1/\Lambda^2)$, instead of the naive expectation of $\mathcal O(1/\Lambda^3)$, so their contributions would be less suppressed than expected.

Another interesting possibility arises in models where the cLFV is mediated by heavy scalars, such as a two Higgs doublet model (2HDM) with off-diagonal Yukawa interactions.
In this scenario, $\ell_i\to\ell_j\gamma$ decays are induced at one-loop level, however they are suppressed by three chiral flips and therefore the two-loop (Barr-Zee diagrams) contributions are actually the dominant ones~\cite{HISANO2011380}. 
On the other hand, the $\ell_i\to\ell_j\gamma\gamma$ decays do not suffer from this chirality suppression, the dominant contributions are at the one-loop and, consequently, they can have ratios comparable to those of $\ell_i\to\ell_j\gamma$.

\begin{figure}[t!]
 \centering
\includegraphics[width=\columnwidth]{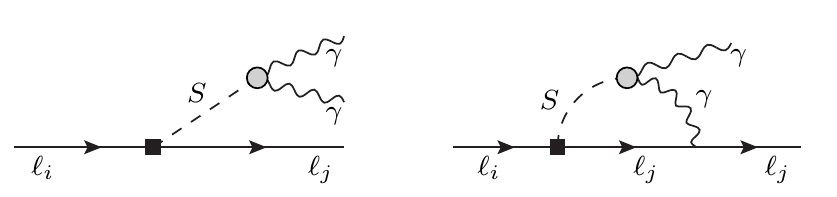}
\caption{Example of diagrams generating $\ell_i \to \ell_j \gamma\gamma$ and $\ell_i \to \ell_j \gamma$ mediated by a  scalar with off-diagonal Yukawa couplings and an effective vertex to two photons.}
\label{diagsHEFT}
\end{figure}

More concretely,  one can consider a scenario containing a heavy scalar $S$ with off-diagonal Yukawa couplings to leptons and an effective vertex to two photons (which matches to the framework in Ref.~\cite{HISANO2011380} when integrating out the top and $W$ boson).
The double and single photon decays are then generated by  diagrams such as those in Fig.~\ref{diagsHEFT}, which in the heavy scalar limit reduce to a local interaction and  Fig.~\ref{fig:diagrams}, respectively.
We have explicitly checked that, in this heavy limit, we recover our EFT result of Eq.~\eqref{BRrelations} with $\Lambda=m_S$.

\section{Conclusions}\label{sec:Conclusions} 

Pursuing an effective field theory approach, we have derived model-independent upper limits on the rates of $\ell_i\to\ell_j\gamma\gamma$ from the current experimental limits on $\ell_i\to\ell_j\gamma$. Our indirect limits are, even under the most conservative assumptions, significantly more stringent than the current direct limits, concretely three orders of magnitude for $\mu\rightarrow e\gamma\gamma$ and four orders of magnitude for $\tau\rightarrow \mu\gamma\gamma$. When $\ell_i\to\ell_j\gamma\gamma$ is dominantly generated by the same dimension-5 operators generating $\ell_i\to\ell_j\gamma$, the stringent limits on the latter preclude the observation of the former in planned experiments. In contrast, in scenarios where the dimension-5 operators are suppressed compared to the dimension-7 operators, the rare decays $\ell_i\to\ell_j\gamma\gamma$ can be enhanced compared to $\ell_i\to\ell_j\gamma$, which in our EFT approach is only generated at the one-loop level. In this class of scenarios, the rare decay $\tau\rightarrow\mu\gamma\gamma$ could be at the reach of the Belle~II experiment or at a proposed Super Tau Charm Facility,  and could constitute an stringent probe of lepton flavor violation. We also discussed some possible UV-complete scenarios where $\ell_i\to\ell_j\gamma\gamma$ is enhanced.


\vspace*{.4cm}
\begin{acknowledgements}
F.F., M.M.~and P.R.~are grateful to Eduard de la Cruz, Iv\'an Heredia and  Alex Miranda for discussions.
This work has been supported by the Collaborative Research Center SFB1258, by the Deutsche Forschungsgemeinschaft (DFG, German Research Foundation) under Germany's Excellence Strategy - EXC-2094 - 390783311, and by the Science and Technology Facilities Council [grant number ST/T004169/1].
F.F.~and M.M.~acknowledge Conacyt scholarships and P.R.~is grateful to funding from `C\'atedras Marcos Moshinsky' (Fundaci\'on Marcos Moshinsky) and  `Paradigmas y Controversias de la Ciencia 2022' (project number 319395,  Conacyt), which supported M.M.
X.M.~acknowledges funding from the European Union’s Horizon Europe Programme under the Marie Sk\l{}odowska-Curie grant agreement no.~101066105-PheNUmenal.
\end{acknowledgements}


\bibliography{biblio}

\onecolumngrid
\appendix
\section{Differential decay rates}\label{sec:Appendix}
In this appendix we present the double differential decay width for the $\ell_i\to\ell_j\gamma\gamma$ processes, including the contributions from effective operators up to dim-7. The differential rate can be cast as:
\begin{align}
\frac{\dd^2\Gamma(\ell_i\to\ell_j\gamma\gamma)}{\dd E_{\gamma} \dd E_{\gamma'}}=
\frac{\dd^2\Gamma(\ell_i\to\ell_j\gamma\gamma)}{\dd E_{\gamma} \dd E_{\gamma'}}\Big|_{\rm dim-5}
+\frac{\dd^2\Gamma(\ell_i\to\ell_j\gamma\gamma)}{\dd E_{\gamma} \dd E_{\gamma'}}\Big|_{\rm dim-7}
+\frac{\dd^2\Gamma(\ell_i\to\ell_j\gamma\gamma)}{\dd E_{\gamma} \dd E_{\gamma'}}\Big|_{\rm int}\,,
\end{align}
where in an obvious notation, dim-5 denotes the contribution from the Lagrangian  Eq.~\eqref{eq:Leffdim5}, dim-7 from the Lagrangian Eq.~\eqref{eq:Leff}, and  int is the interference term. Explicitly, and neglecting the mass of the lepton in the final state, we obtain:
\begin{align}
\frac{\dd^2\Gamma(\ell_i\to\ell_j\gamma\gamma)}{\dd E_{\gamma} \dd E_{\gamma'}}\Big|_{\rm dim-7} &=
\frac{\big|G_{ij}\big|^2}{16\pi^3}\, m_i^2 \big(m_i-E_\gamma-E_{\gamma'}\big)\big(m_i-2(E_\gamma+E_{\gamma'})\big)^2\,, \\
\frac{\dd^2\Gamma(\ell_i\to\ell_j\gamma\gamma)}{\dd E_{\gamma} \dd E_{\gamma'}}\Bigg|_{\rm dim-5}
&=
\frac{\alpha \left(|D^{ij}_{R}|^2+|D^{ij}_{L}|^2\right) }{ 4 E_\gamma E_{\gamma'}+m_i^2-2 m_i (E_\gamma+E_{\gamma'})}
\frac{m_i-2 (E_\gamma+E_{\gamma'})}{4 \pi ^2 E_\gamma^2 E_{\gamma'}^2}
\bigg\{48 E_\gamma^3 E_{\gamma'}^3 -m_i^4 \big(E_\gamma-E_{\gamma'}\big)^2  \nonumber\\
&
+2 E_\gamma E_{\gamma'} m_i^2\left(E_\gamma^2+6 E_\gamma E_{\gamma'}+E_{\gamma'}^2\right)
-E_\gamma E_{\gamma'} m_i (E_\gamma+E_{\gamma'})\left(24 E_\gamma E_{\gamma'}+m_i^2\right)\bigg\}\,, \\
\frac{\dd^2\Gamma(\ell_i\to\ell_j\gamma\gamma)}{\dd E_{\gamma} \dd E_{\gamma'}}\Bigg|_{\rm int} 
&
= -\frac{e\,m_i\big(m_i-2(E_\gamma+E_{\gamma'})\big)^2}{4 \pi^3} \Re\bigg\{ 
D^{ij*}_L \left( {G}^{ij}_{SL}+i \tilde{G}^{ij}_{SL} \right)  
+ D^{ij*}_R  \left(G^{ij}_{SR} - i \tilde{G}^{ij}_{SR} \right)
\bigg\} \,,
\end{align}
with kinematical ranges for the photon energies $0\leq E_\gamma \leq m_i/2$, $m_i/2-E_\gamma\leq E_{\gamma'} \leq m_i/2$. 
Notice that the dimension-5 contribution suffers from both infrared and collinear singularities, which we can avoid by introducing a regulator such that $E_\gamma^{\rm cut} \leq E_\gamma \leq m_i/2-E_\gamma^{\rm cut}$, $m/2-E_\gamma\leq E_{\gamma'} \leq m_i/2-E_\gamma^{\rm cut}$.
Strictly, one should introduce different regulators, however we assume them to be the same for simplicity.

\end{document}